\documentclass{iaus}
\usepackage{graphicx,natbib}

\newcommand{\msun}{M_{\odot}}
\newcommand{\sfrff}{\mbox{SFR}_{\rm ff}}
\newcommand{\calm}{\mathcal{M}}
\newcommand{\avir}{\alpha_{\rm vir}}

\newcommand{\aap}{\textit{A\&A}}
\newcommand{\apj}{\textit{ApJ}}
\newcommand{\apjl}{\textit{ApJL}}
\newcommand{\apjs}{\textit{ApJS}}

\newcommand{\mnras}{\textit{MNRAS}}

\title[Turbulence, Feedback, and Slow Star Formation] %% give here short title %%
{Turbulence, Feedback, and Slow Star Formation}

\author[Krumholz]   %% give here short author list %%
{Mark R. Krumholz$^1$}

\affiliation{$^1$Department of Astrophysical Science, Princeton University, Peyton Hall, Ivy Lane, Princeton, NJ 08544, USA \break email: krumholz@astro.princeton.edu}

\pubyear{2006}
\volume{237}  %% insert here IAU Symposium No.
\pagerange{}
\date{}
\jname{Triggered Star Formation in a Turbulent ISM}
\editors{B.~G. Elmegreen and J. Palous, eds.}
\begin{document}

\maketitle

\begin{abstract}
One of the outstanding puzzles about star formation is why it proceeds so slowly. Giant molecular clouds convert only a few percent of their gas into stars per free-fall time, and recent observations show that this low star formation rate is essentially constant over a range of scales from individual cluster-forming molecular clumps in the Milky Way to entire starburst galaxies. This striking result is perhaps the most basic fact that any theory of star formation must explain. I argue that a model in which star formation occurs in virialized structures at a rate regulated by supersonic turbulence can explain this observation. The turbulence in turn is driven by star formation feedback, which injects energy to offset radiation from isothermal shocks and keeps star-forming structures from wandering too far from virial balance. This model is able to reproduce observational results covering a wide range of scales, from the formation times of young clusters to the extragalactic IR-HCN correlation, and makes additional quantitative predictions that will be testable in the next few years.
\end{abstract}

\section{Introduction}

\citet{zuckerman74} were the first to point out perhaps the most surprising fact about star formation: it is remarkably slow. Inside the solar circle there are roughly $M_{\rm mol}\approx 10^9$ $\msun$ of molecular gas \citep{bronfman00}, organized into giant molecular clouds (GMCs) with typical densities of $\sim 100$ H atoms cm$^{-3}$, giving a free-fall time of about $t_{\rm ff}\approx 4$ Myr \citep{mckee99a}. However, the star formation rate in the Milky Way is only $\sim 3$ $\msun$ yr$^{-1}$ \citep{mckee97}, vastly less than the rate of $\sim 250$ $\msun$ yr$^{-1}$ that one would expect if molecular clouds were converting their mass into stars on a free-fall time scale. More recent observations of nearby Milky Way-like galaxies \citep{wong02} find that this factor of $\sim 100$ discrepancy occurs in them too. Nor is the discrepancy any smaller in systems like ULIRGs with much larger star formation rates. For example, \citet{downes98} find that Arp 220 contains roughly $2\times 10^9$ $\msun$ of molecular gas with a typical free-fall time of $\sim 0.5$ Myr, but the observed star formation rate of $\sim 50$ $\msun$ yr$^{-1}$ is a factor of 100 smaller than $M_{\rm mol}/t_{\rm ff}$.

\begin{figure}
\centerline{
\includegraphics{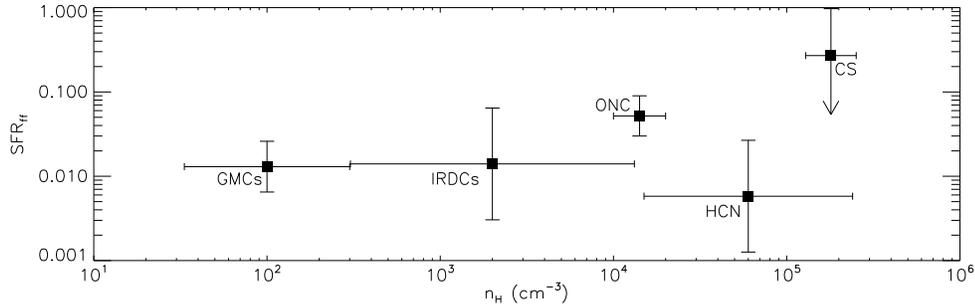}
}
\caption{\label{sfrffn}
Fraction of mass converted into stars per free-fall time ($\sfrff$) versus characteristic density for various objects. We show GMCs, infrared dark clouds (IRDCs), HCN-emitting gas clumps, the Orion Nebula Cluster, and CS-emitting gas clumps. Note that the CS point is only an upper limit. The Figure is adapted from \citet{krumholz06c}.
}
\end{figure}

Recently, \citet{krumholz06c} pointed out that objects much denser than GMCs form stars just as slowly. If one repeats the \citet{zuckerman74} calculation of dividing total mass by characteristic free-fall time for any class of dense, gaseous objects (e.g.\ infrared dark clouds, dense molecular clumps), one again obtains a rate roughly 100 times larger than the observed star formation rate. This is true in galaxies from normal spirals to ULIRGs, and for objects with densities from $\sim 100$ cm$^{-3}$, typical of GMCs, to $\sim 10^4-10^5$ cm$^{-3}$, typical of molecular clumps forming rich star clusters. The trend may continue to even higher densities. Figure \ref{sfrffn} summarizes the observations.

The apparent universality of this factor of 100 discrepancy over such an immense range of densities and galactic environments suggests it must be set by fundamental physics. In these proceedings I present a model that attempts to explain the observations as a result of the properties of supersonic turbulence, and in which that turbulence is itself a side-effect of the star formation process. This theory has two parts: a physical mechanism by which turbulence determines the factor of 100, described in \S~\ref{turbregulation}, and a physical mechanism for generating the turbulence, described in \S~\ref{sfregulation}.

\section{Turbulent Regulation of the Star Formation Rate}
\label{turbregulation}

\citet{krumholz05c} propose a simple model for how turbulence regulates the star formation rate, based on the premise that star formation occurs in any sub-region of a molecular cloud in which the gravitational potential energy exceeds the kinetic energy in turbulent motions. This is sufficient to determine the star formation rate in a supersonically turbulent isothermal medium, because such media are governed by two universal properties. First, they obey a linewidth-size relation, meaning that the velocity dispersion over a region of size $\ell$ varies as roughly $\ell^{1/2}$ \citep{larson81}. Second, they show a lognormal distribution of densities \citep{padoan02}. 

These determine the star formation rate as follows: the linewidth-size relation sets the kinetic energy per unit mass in any given sub-region of a cloud, normalized to the cloud's total kinetic energy. From this, one can show that the potential energy will be larger than the kinetic energy in regions where the density exceeds a certain critical value. In turn, the density probability distribution determines what fraction of the mass is at densities larger than this critical value. Bound regions collapse on a free-fall time scale. This determines the dimensionless star formation rate $\sfrff$, defined as the fraction of its mass that a gas cloud turns into stars per mean-density free-fall time, in terms of two dimensionless numbers: the cloud's Mach number $\calm$ and virial ratio $\avir$ (roughly its ratio of kinetic to potential energy). For $\avir$ and $\calm$ in the range observed for real clouds, one may approximate the relationship by a powerlaw
\begin{equation}
\label{sfrffeqn}
\sfrff \approx 0.073\avir^{-0.68} \calm^{-0.32}.
\end{equation}

This is an extremely powerful result, and it allows numerous comparisons to observation. First, notice that, due to its very weak dependence on $\calm$, $\sfrff$ is a few percent in any virialized, supersonically turbulent object, regardless of its density or environment. This is exactly what the observations summarized in Figure 1 demand; a corollary is that this model naturally explains the extra-Galactic IR-HCN correlation \citep{gao04b, krumholz06c}. A second prediction of this model is that achieving the star formation efficiencies of tens of percent estimated for rich star clusters requires that star formation continue for several crossing times, a result in good agreement with the observed age spreads of young clusters \citep{tan06a}.

A third test of the model is to use it to predict the star formation rate in the Milky Way from equation (\ref{sfrffeqn}) and the observed  properties of Milky Way molecular clouds. Doing so gives a predicted star formation rate of 5.3 $\msun$ yr$^{-1}$ in the Milky Way, within a factor of 2 of the observed value \citep{krumholz05c}. Moreover, observations of the molecular cloud populations of nearby galaxies such as M33, M64, and the LMC \citep[see review by][]{blitz06a} are starting to reach precisions comparable to those available for the Milky Way. Repeating this calculation for these cloud populations and comparing to observed star formation rates provides a direct future observational test of this model. 

If one adopts the additional hypotheses that molecular clouds in a galaxy should have masses of $\sim 1$ Jeans mass in the galactic disk, and should be in rough pressure balance with the rest of the ISM, then one may extend the model to predict the star formation rate in galaxies as a function of their gas surface densities, rotation rates, and molecular fractions. This prediction agrees with the data of \citet{kennicutt98a} as well as Kennicutt's purely empirical fit. Recent observations show that the model also agrees with the radial distribution of star formation in the Milky Way \citep{luna06}.

\section{Star Formation Regulation of the Turbulence}
\label{sfregulation}

\begin{figure}
\centerline{
\includegraphics{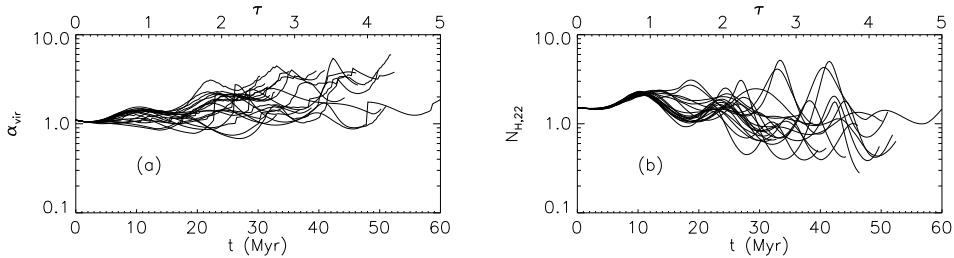}
}
\caption{\label{gmcmodels}
Virial ratio (panel a) and column density in units of $10^{22}$ cm$^{-2}$ (panel b) versus time measured in years ($t$) and measured in cloud crossing times ($\tau$) for a sample of GMC evolution models. Lines end when clouds are disrupted by HII regions. The Figure is adapted from \citet{krumholz06d}.
}
\end{figure}

Having shown that turbulent regulation very naturally explains a large number of observations about star formation, we now turn to the question of the origin of the turbulence itself. This is a problem because simulations \citep[e.g.][]{stone98} show that, if it is not continually driven, supersonic turbulence will decay away over time scales much shorter than the observationally-estimated $\sim 30$ Myr lifetimes of GMCs in local group galaxies \citep{blitz06a}. Since the clouds all have virial ratios $\sim 1$ and roughly constant column densities $\sim 10^{22}$ cm$^{-2}$, something must be driving the turbulence and keeping the clouds in at least approximate equilibrium.

To investigate this problem, \citet{krumholz06d} construct simple one-dimensional semi-analytic models of GMCs, the goal of which is to investigate their global energy balance. In these models one follows the evolution of GMCs using the non-equilibrium virial and energy conservation equations for a homologously-moving, evaporating cloud, including source terms describing decay of turbulence at the rates measured by simulations, and a countervailing injection of energy (and removal of mass) by star formation feedback. Both energy injection and mass loss are dominated by HII regions launched by newborn star clusters \citep{matzner02}, so the models focus on them.

These models show that feedback is able to explain the observed properties of GMCs extremely well. Feedback destroys giant clouds in $\sim 30$ Myr, during which time they turn $5-10\%$ of their mass into stars, and remain turbulent, virialized, and at constant column densities. Figure \ref{gmcmodels} shows some typical examples of molecular cloud evolution, which are in good agreement with observations. In contrast, clouds with masses $\sim 10^4$ $\msun$, like those in the solar neighborhood, survive only $\sim 1$ crossing time. The next step in this sort of modeling is to do full ionizing radiative-trainsfer MHD simulations to study feedback in a more realistic context, a project already underway \citep{krumholz06e}.

\section{Summary}

Feedback-driven turbulence provides a simple, natural explanation for a host of observations about star formation. The quantitative prediction of the star formation rate one derives by computing the fraction of bound mass in a turbulent, virialized object, using no physics other than the invariant properties of supersonic isothermal turbulence, matches observations of the rate of star formation in dense gas, the rate and radial distribution of star formation in the Milky Way, the Kennicutt law, and the extragalactic IR-HCN correlation. The turbulence is in turn driven by the feedback from star formation itself. The physics of the driving process explains the observed lifetimes, column densities, and virial ratios of giant molecular clouds in local group galaxies.

%\bibliographystyle{apj}
%\bibliography{refs}

\begin{acknowledgments}
I thank T.~A. Gardiner, C.~D. Matzner, C.~F. McKee, J.~M. Stone, and J.~C. Tan, all of whom collaborated with me on parts of this project. I acknowledge support from NASA through Hubble Fellowship grant \#HSF-HF-01186 awarded by STScI, which is operated by AURA for
NASA, under contract NAS 5-26555.
\end{acknowledgments}

\end{document}